\documentclass[prl,twocolumn,showpacs,floatfix,superscriptaddress]{revtex4-1}

\usepackage{amssymb,amsmath,graphicx}

\begin{document}

\title{Spin-stretching modes in anisotropic magnets: spin-wave excitations in the multiferroic Ba$_2$CoGe$_2$O$_7$}

%
% authors and affiliations
%
\author{K. Penc}
\affiliation{Institute for Solid State Physics and Optics, Wigner Research
Centre for Physics, Hungarian Academy of Sciences, H-1525 Budapest, P.O.B. 49, Hungary}
\affiliation{Department of Physics, Budapest University of
Technology and Economics and Condensed Matter Research Group of the Hungarian Academy of Sciences, 1111 Budapest, Hungary}
\author{J. Romh\'anyi}
\affiliation{Institute for Solid State Physics and Optics, Wigner Research
Centre for Physics, Hungarian Academy of Sciences, H-1525 Budapest, P.O.B. 49, Hungary}
\affiliation{Department of Physics, Budapest University of
Technology and Economics and Condensed Matter Research Group of the Hungarian Academy of Sciences, 1111 Budapest, Hungary}
\author{T. R\~o\~om}
\affiliation{National Institute of Chemical Physics and Biophysics, 12618 Tallinn, Estonia}
\author{U. Nagel}
\affiliation{National Institute of Chemical Physics and Biophysics, 12618 Tallinn, Estonia}
\author{\'A. Antal}
\affiliation{Department of Physics, Budapest University of
Technology and Economics and Condensed Matter Research Group of the Hungarian Academy of Sciences, 1111 Budapest, Hungary}
\author{T. Feh\'er}
\affiliation{Department of Physics, Budapest University of
Technology and Economics and Condensed Matter Research Group of the Hungarian Academy of Sciences, 1111 Budapest, Hungary}
\author{A. J\'anossy}
\affiliation{Department of Physics, Budapest University of
Technology and Economics and Condensed Matter Research Group of the Hungarian Academy of Sciences, 1111 Budapest, Hungary}
\author{H. Engelkamp}
\affiliation{High Field Magnet Laboratory, Institute for Molecules and Materials, Radboud University,
6525 ED Nijmegen, The Netherlands}
\author{H. Murakawa}
\affiliation{Multiferroics Project, ERATO, Japan Science and Technology Agency (JST), Japan c/o The University of Tokyo, Tokyo 113-8656, Japan}
\affiliation{Quantum-Phase Electronics Center, Department of Applied Physics,
The University of Tokyo, Tokyo 113-8656, Japan}
\author{Y. Tokura}
\affiliation{Multiferroics Project, ERATO, Japan Science and
Technology Agency (JST), Japan c/o The University of Tokyo, Tokyo
113-8656, Japan} \affiliation{Quantum-Phase Electronics Center,
Department of Applied Physics, The University of Tokyo, Tokyo
113-8656, Japan} \affiliation{Department of Applied Physics, The
University of Tokyo, Tokyo 113-8656, Japan}
\affiliation{Cross-correlated materials group (CMRG) and correlation
electron research group (CERG), RIKEN Advanced Science Institute,
Wako 351-0198, Japan}
\author{D. Szaller}
\affiliation{Department of Physics, Budapest University of
Technology and Economics and Condensed Matter Research Group of the Hungarian Academy of Sciences, 1111 Budapest, Hungary}
\author{S. Bord\'acs}
\affiliation{Department of Physics, Budapest University of
Technology and Economics and Condensed Matter Research Group of the Hungarian Academy of Sciences, 1111 Budapest, Hungary}
\affiliation{Multiferroics Project, ERATO, Japan Science and Technology Agency (JST), Japan c/o The University of Tokyo, Tokyo 113-8656, Japan}
\affiliation{Quantum-Phase Electronics Center, Department of Applied Physics,
The University of Tokyo, Tokyo 113-8656, Japan}
\author{I. K\'ezsm\'arki}
\affiliation{Department of Physics, Budapest University of
Technology and Economics and Condensed Matter Research Group of the Hungarian Academy of Sciences, 1111 Budapest, Hungary}
\affiliation{Multiferroics Project, ERATO, Japan Science and Technology Agency (JST), Japan c/o The University of Tokyo, Tokyo 113-8656, Japan}

\date{\today}
\begin{abstract}
We studied spin excitations of the non-centrosymmetric
Ba$_2$CoGe$_2$O$_7$ in high magnetic fields up to 33\,T. In the
electron spin resonance and far infrared absorption spectra we found
several spin excitations beyond the two conventional magnon modes
expected for such a two-sublattice antiferromagnet. We show that a
multi-boson spin-wave theory describes these unconventional modes,
including spin-stretching modes, characterized by oscillating
magnetic dipole and quadrupole moment.
The lack of the inversion symmetry allows each
mode to become electric dipole active. We expect that the
spin-stretching modes can be generally observed in inelastic neutron
scattering and light absorption experiments in a broad class of
ordered $S>1/2$ spin systems with strong single-ion anisotropy
and/or non-centrosymmetric lattice structure.
\end{abstract}
%
% PACS numbers
%
\maketitle

Magnons are collective spin excitations in crystals with long-range
magnetic order, often investigated by electromagnetic absorption and
neutron scattering experiments.
Both classical and quantum spin-wave theory of $S=1/2$ systems predict
one magnon branch in the spin-excitation spectrum for each spin in the
magnetic unit cell \cite{VanKranendonk1958}.
This rule about the number of magnon branches is generally accepted
and experimentally verified for $S>1/2$ spin systems as long as the conventional
spin-wave theory applies, requiring that the lengths (i.e., the
absolute values of the expectation values) of the spins are preserved in
the excited states and only their orientations change relative to the
ground-state configuration \cite{Turov}.  
However, the picture of one
magnon mode per spin in the magnetic unit cell needed to be surpassed in
several $f$-electron compounds with complicated quadrupolar ordering,
such as CeB$_6$ \cite{Shiina2003} and UO$_2$ \cite{Carretta2010}.

Recently, additional spin-wave modes have been observed by far infrared
(FIR) spectroscopy \cite{Kezsmarki2011} and inelastic neutron scattering
(INS) \footnote{C. de la Cruz, private communication}
in Ba$_2$CoGe$_2$O$_7$, a simple two-sublattice easy-plane antiferromagnet
(AF) with $S=3/2$ spins \cite{Zheludev2003,Murakawa2010}.
This material has attracted much interest owing to its multiferroic
ground state where delicate magnetic control of the ferroelectric
polarization \cite{Yi2008,Murakawa2010} and chirality \cite{Bordacs}
were realized.
Moreover, spin waves in Ba$_2$CoGe$_2$O$_7$ exhibits giant
directional dichroism and natural optical activity at THz
frequencies due to the large ac magnetoelectric effect \cite{Kezsmarki2011,Bordacs}.
A recent numerical diagonalization study on finite spin clusters
found, besides the two conventional AF modes, additional
spin resonances with peculiar optical properties
\cite{Bordacs,Miyahara2011}.
Nevertheless, the understanding of the unconventional magnon modes and
the coupled dynamics of spins and electronic polarization on a
fundamental level remained an open issue.

\begin{figure*}[t!]
\includegraphics[width=6.5in]{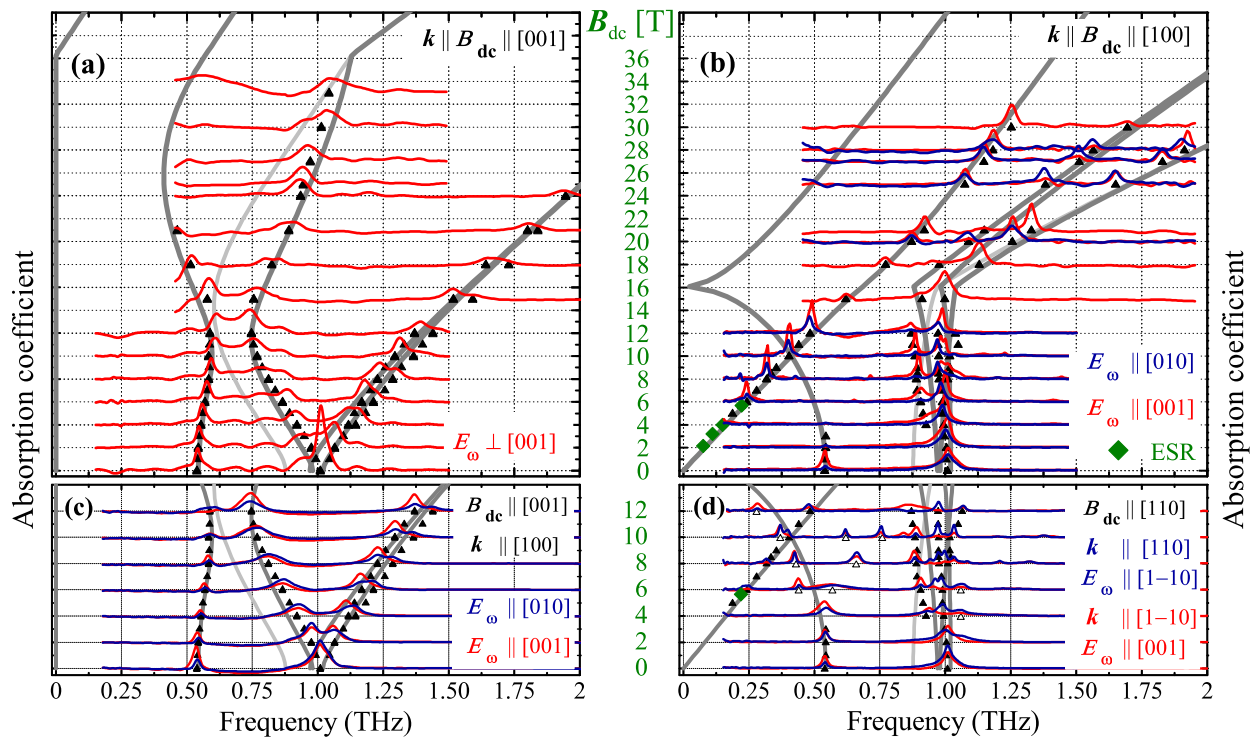}
\caption{(color online). Magnetic field dependence of the absorption
spectra in Ba$_2$CoGe$_2$O$_7$ below 2 THz for a representative set
of light polarizations. The spectra are shifted vertically
proportional to the magnitude of the field, $B_{\rm dc}$. The
distance between horizontal grid lines corresponds to 20\,cm$^{-1}$
in panels (a) and (c), and 30\,cm$^{-1}$ in (b) and (d). The
direction of $B_{\rm dc}$ is indicated in each panel and the spectra
for different polarizations and propagation directions (${\mathbf
k}$) of light are distinguished by the color. Black triangles and
green diamonds represent the position of the resonances determined
from the FIR and ESR spectra, respectively. The grey lines show the
field dependence of the modes obtained in our multi-boson spin-wave
approach. (d) For $B_{\rm dc}\perp[001]$ in some polarization
configurations we observed additional modes (open triangles) that
are not explained by the theory. Two cases ($E\|[1\bar10]$ and $E\|[001]$) are shown here.}
\label{fig:spectra}
\end{figure*}

In this Letter, we investigate the spin-wave excitations in
Ba$_2$CoGe$_2$O$_7$ over a broad photon energy range combining
electron spin resonance (ESR) and high-resolution FIR spectroscopy.
The largest magnetic field, 33\,T, applied in this study
drastically changes the antiferromagnetic spin configuration for any
field direction, in contrast to former experiments restricted to $B_{\rm
dc}\leq12\rm\,T$.
The orientation of $B_{\rm dc}$ and the light polarization relative
to the main crystallographic axes were systematically varied in
order to map the field dependence and the selection rules of the
modes.
We derive a multi-boson spin-wave theory and show that a large single-ion
anisotropy plays a key role in the emergence of new magnetic
excitations involving the oscillation of spin length, and that the lack of
inversion symmetry, a necessary condition for the dc and ac
magnetoelectric effects, renders these spin-waves  electric-dipole active.

Ba$_2$CoGe$_2$O$_7$ has a non-centrosymmetric tetragonal space group,
P$\overline{4}2_1m$. The magnetic Co$^{2+}$ ions are surrounded by
tetrahedra of oxygens compressed along the [001] tetragonal axis.
Due to the lack of inversion symmetry a coupling between spins and
local polarization appears \cite{Toledano2011,Yamauchi2011}.
This was observed as a magnetic-order induced ferroelectricity in
this family of materials including Ba$_2$CuGe$_2$O$_7$
\cite{Yi2008,Murakawa2010}, Ca$_x$Sr$_{2-x}$CoSi$_2$O$_7$
\cite{Akai2009,*Akai2010}, and Ba$_2$MnGe$_2$O$_7$
\footnote{H. Murakawa, private communication}.

Here we study spin-wave resonances of Ba$_2$CoGe$_2$O$_7$ on
high-quality single crystals \cite{Murakawa2010} in the magnetic
phase at $T=3.5\rm\,K$. ESR spectroscopy was performed at 75, 111, 150,
and 222\,GHz using solid state oscillators, while FIR transmission was
measured by Fourier transform spectroscopy over the region of
0.15--2\,THz (0.6--8\,meV) with a resolution of 15\,GHz.

An overview of the spectra for representative directions of the magnetic
field is given in Fig.~\ref{fig:spectra}. Two sharp peaks are present in
the zero-field FIR absorption spectra at $f\sim0.5$ and 1\,THz, in
accordance with former studies \cite{Kezsmarki2011,Bordacs}. The first
is assigned to the usual optical magnon branch gapped by magnetic
anisotropy of mostly single-ion origin. The second is not captured by
conventional spin-wave theory, and was shown to respond to both the
magnetic and electric component of light and termed as an electromagnon
\cite{Kezsmarki2011}. For $B_{\rm dc}$ along the tetragonal axis as in
Fig.~\ref{fig:spectra}(a) and (c), the 1\,THz mode shows a V-shape
splitting with a double-peak structure on the high-energy side. The
double-peak structure is clearly visible when $E_{\omega}\parallel[010]$
[see the blue curves in Fig.~\ref{fig:spectra}(c)]. The frequency of the
0.5\,THz mode slightly increases in low fields
[Fig.~\ref{fig:spectra}(a) and (c)], however, it turns back after an
avoided crossing with the lower branch of the 1\,THz resonance at
$B_{\rm dc}\approx12\rm\,T$.

The rotation of $B_{\rm dc}$ from the tetragonal axis to the tetragonal
plane affects all the modes drastically (Fig.~\ref{fig:spectra}). For $B_{\rm
dc}\parallel[100]$, the 1\,THz mode is again split into three distinct
lines. However, they exhibit only a weak softening up to a kink at
$B_{\rm dc}\approx16\rm\,T$, from where the resonance frequencies start
to increase quickly [Fig.~\ref{fig:spectra}(b)]. The magnon mode at
$f\sim0.5\rm\,THz$ becomes silent with increasing field for both
polarization directions in the Faraday geometry (${\mathbf k}||B_{\rm
dc}$).

An additional low-frequency mode appears in the ESR and FIR spectra
when $B_{\rm dc}$ is within the tetragonal plane, breaking the fourfold
rotoinversion symmetry of the lattice. This mode corresponds to the
quasi-Goldstone mode of an easy-plane AF when $B_{\rm dc}=0$. The
frequency of this mode is not affected measurably by the orientation of
$B_{\rm dc}$ in the plane and follows a linear field dependence down to
75\,GHz. Hence, the in-plane anisotropy gap is less than $75\rm\,GHz$.

If $B_{\rm dc}$ is in the tetragonal plane, the number of
observed resonances exceeds six in some polarization configurations.
Two representative cases are presented in Fig.~\ref{fig:spectra}(d).
The 0.5\,THz mode suddenly splits into a sharp and a broad feature
at $B_{\rm dc}=5\rm\,T$, while the 1\,THz branch consists of at
least four resonances. At $\gtrsim12\rm\,T$, the number of modes is
reduced.

As a microscopic model, we consider the Hamiltonian below to describe
the $S = 3/2$ spin Co$^{2+}$ ions. Following Ref.~\onlinecite{Miyahara2011},
we have a large single-ion anisotropy $\Lambda$, but we introduce an
anisotropic exchange coupling ($J$ and $J_z$) and neglect the
Dzyaloshinskii-Moriya term that appeared to have negligible effect on the
excitations:
\begin{eqnarray}
\mathcal{H}&=&J\sum_{\langle i,j\rangle}\left(S^x_i  S^x_j+ S^y_i  S^y_j\right)+J_z \sum_{\langle i,j\rangle} S^z_i  S^z_j+
\nonumber\\&&
\sum_i \left[\Lambda \left(S^z_i\right)^2 + g_{zz} h_z S^z_i + g_{xx} (h_x S^x_i + h_y S^y_i)\right],
\label{eq:Hamiltonian_1}
\end{eqnarray}
where $\langle i,j\rangle$ indicates nearest neighbor pairs, and
the $x$, $y$, and $z$ axes are parallel to the $[110]$, $[1\bar
10]$, and $[001]$ crystallographic directions, respectively. $g_{xx}=g_{yy}$ and $g_{zz}$ are the principal values of the
$g$ tensor, and $h_\alpha=\mu_{\rm B}B_{\rm dc,\alpha}$ are the components of
the magnetic field.

We assume a site-factorized variational wave function
$|\Psi_0\rangle = \prod_{i\in A} |\Psi_A(i)\rangle  \prod_{i\in B}
|\Psi_B(i)\rangle $  to describe the long-range ordered ground state of two spin sublattices $A$ and $B$ ($i$ is the site index).
For example, when $B_{\rm dc}\parallel[110]$ and $J_z/J\lessapprox 4$,
we get a canted N\'eel state \cite{PhysRevB.84.184427,*PhysRevB.84.224419} where the expectation values of spin components are
\begin{equation}
\langle \Psi_X(i) | \hat {\mathbf S}_X |\Psi_X(i) \rangle  = \frac{3 \eta (\eta +1)}{3 \eta^2+1} (\cos \varphi_X ,\sin \varphi_X ,0)\,.
\label{eq:spinave}
\end{equation}
The two variational parameters $\eta$ and
$\varphi_A=-\varphi_B$ are determined from the minimization
of the energy ($X=A,B$).
We note that $\eta\neq 1$ corresponds to a spin with length smaller than 3/2,
the consequence of the on-site anisotropy.

This $|\Psi_0\rangle$ serves as a
starting point to study excitations: we introduce four orthogonal
bosons on each site, denoted by $a^\dagger_{\nu,X}(i)$, where
$\nu=0,\dots,3$, so that
the variational ground state is $|\Psi_X(i)\rangle =
a^{\dagger}_{0,X}(i) |\text{vacuum}\rangle$.
Any product of the operators on a site can be
expressed as a quadratic form of the four $a$ bosons and they
satisfying the expected commutation
relations. The number of bosons on each site is conserved,
$\sum_{\nu=0}^3 a^{\dagger}_{\nu,X} a^{\phantom{\dagger}}_{\nu,X} = M$,
and $M=1$ for the $S=3/2$ spin.
The linear flavor-wave theory is a $1/M$ expansion, where the
$a^\dagger_{\nu,A}$ and $a^\dagger_{\nu,B}$ with $\nu = 1,2,3$ play
the role of the Holstein-Primakoff bosons and describe the
excitations, the generalized spin-waves. Replacing
$a^\dagger_{0,X}$ and $a^{\phantom{\dagger}}_{0,X}$ with
$(M-\sum_{\nu=1}^3 a^\dagger_{\nu,X}
a^{\phantom{\dagger}}_{\nu,X})^{1/2}$ and performing the expansion
in $1/M$ one can follow the procedure of the conventional spin-wave
theory, and we get a Hamiltonian that is quadratic in
boson operators and straightforward to diagonalize
\footnote{See supplementary material at [URL] for the exact form of the
transformed Hamiltonian and the details of the calculation.}.
A similar approach has been used to
describe, e.g., CeB$_6$ \cite{Shiina2003}, 
the SU(4) Heisenberg model\cite{PhysRevB.60.6584}, 
the ${\mathrm{TlCuCl}}_{3}$ spin ladder\cite{PhysRevB.69.054423}, 
and multipolar excitations in the spin-1\cite{N1984281,*N1988367}
and spin-3/2 \cite{ISI:A1985AXC5300038, *0953-8984-2-6-018} Heisenberg models.

%%%%%%%%%%%%% BEGIN FIG %%%%%%%%%%%%%%%%%%%%%%
\begin{figure}[tb]
\begin{center}
\includegraphics[width=8.5truecm]{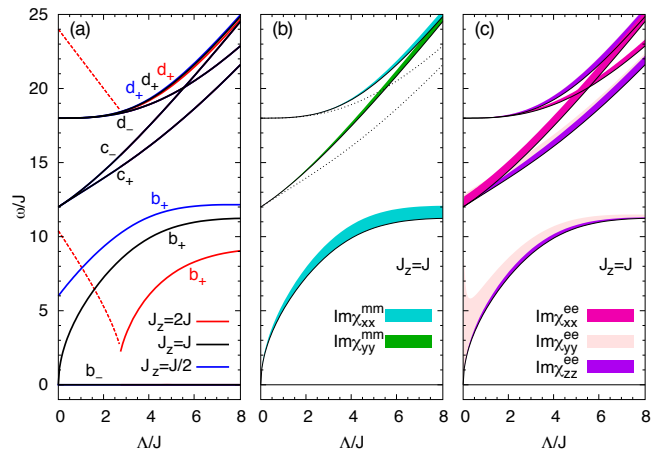}
\caption{(color online). 
(a) The energy of the modes for different
values of $J_z/J$ in zero field. $b_-$ denotes the $\omega=0$
Goldstone mode. Only the $b_+$ and $d_+$ modes depend on $J_z/J$.
The dashed lines indicate modes in the easy-axis AF state that forms
below $\Lambda \approx 2.7 J$ for $J_z=2$.
(b) and (c) shows the imaginary part of the magnetic
$\chi_{\xi\xi}^{mm}(\omega)$ and electric $\chi_{\xi\xi}^{ee}(\omega)$
dynamic susceptibilities, respectively, for $J=J_z$. The shading above
the lines represent the strength of the magnetic and electric response.
}
\label{fig:modes_omega}
\end{center}
\end{figure}
%%%%%%%%%%%%% END FIG %%%%%%%%%%%%%%%%%%%%%%

\begin{figure*}[htb]
\includegraphics[width=6in]{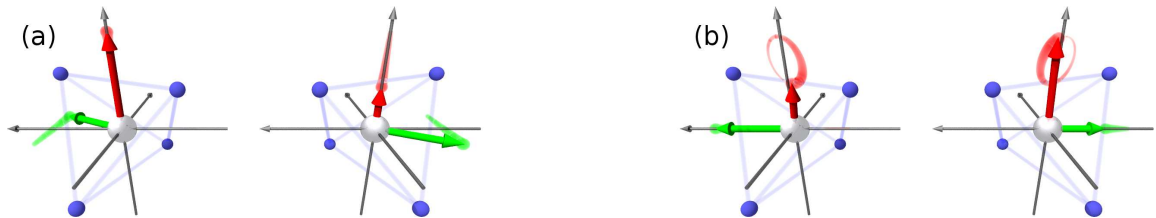}
\caption{(color online). 
Motion of the magnetizations (green arrows) and
the local electric polarizations (red arrows) in the two sublattices (a)
for the Goldstone mode ($b_-$ in Fig.~\ref{fig:modes_omega}) and (b) for the $c_-$ stretching mode. The blue
spheres are the oxygens forming tetrahedral cages around the central Co
ions. The vertical axis is the tetragonal one, while the horizontal axes
point along $[110]$ and $[1\bar 10]$. Apparent tilting of the axes comes
from the perspective view. 
}
\label{fig:modes_anim}
\end{figure*}

The spectrum as a function of $\Lambda/J$ in zero magnetic field and for zero momentum
is shown in Fig.~\ref{fig:modes_omega}(a). It consists of six modes, three for each
 Co spin in the unit cell. A finite anisotropy reduces the O(3) symmetry of the Hamiltonian to O(2),
decreasing the number of zero energy Goldstone modes from two to one.

Let us begin from $\Lambda=0$. Then $\eta = 1$ and the
$a^{\dagger}_{0,X}$ creates a spin coherent state with maximal spin
length 3/2. In this limit the $b_\pm$ branches correspond to
magnons of the standard spin-wave theory and they are decoupled from
the other modes.
The $c_\pm$ and $d_\pm$ are local magnetic transitions with $\Delta
S_W = 2$ and $3$ corresponding to Zeemann energies $12J$ and $18J$,
respectively, in the Weiss field $4\times(3/2)J$ of the neighboring
spins ($S_W$ is the spin component parallel to the Weiss field). The
$c_\pm$ and $d_\pm$ modes are generally silent in neutron, ESR,
and FIR spectra, as the magnetic dipolar matrix elements vanish in the
imaginary part of the dynamic magnetic susceptibility, $\text{Im}
\chi^{mm}_{\alpha\alpha}(\omega) \propto \sum_f |\langle
f|S^\alpha|\Psi_0\rangle|^2 \delta(\omega-\omega_f+\omega_0)$. These
transitions can only be excited by quadrupolar or higher order spin
operators.

As we turn on $\Lambda>0$, $\eta$ increases and the spin length decreases in
the N\'eel ground state  [Eq.~(\ref{eq:spinave})]. 
The modes labeled as $c_-$ and $c_+$ in Fig.~\ref{fig:modes_omega}
are spin-stretching modes, with spin length oscillating in and out of phase on the two sublattices, respectively. Hence  
 $c_-$ is excited by the $S^y$ spin operator, with
a finite weight in $\text{Im} \chi^{mm}_{yy}(\omega)$
that vanishes as $(\Lambda/J)^2$ when $\Lambda/J \to 0$.
Most of the weight
in $\text{Im} \chi_{xx}^{mm}(\omega)$ comes from the low-energy
$b_+$ mode, while the contribution of $d_+$ to $\text{Im} \chi_{xx}^{mm}(\omega)$
is  $\propto (\Lambda/J)^4$,
so that the sum rule  $\int\!\! d\omega \text{Im}
\chi_{xx}^{mm}(\omega)/\omega = g^2_{xx}/8J$ is fulfilled.
$\text{Im} \chi_{zz}^{mm}(\omega)$ is zero for all but the Goldstone
mode $b_-$.

For a large on-site anisotropy ($\Lambda\gg J,J_z$), $ \eta \to
\Lambda/3J$ in the leading order and the $S^z=\pm 3/2$ states are
suppressed in the ground state, reducing the spin length to 1. We
recover the spectra of isolated spins with single-ion anisotropy: two
modes with energies $\omega/\Lambda \to 0$ and four modes with
$\omega \to 2\Lambda$, in agreement with Ref.~\onlinecite{Miyahara2011}.

From the  analysis of the dynamic magnetic susceptibility it follows that these unconventional spin
excitations become observable by ESR, FIR, and neutron
scattering as soon as the single-ion anisotropy gets significant.
Moreover, if the crystal lattice breaks the inversion symmetry, spin
quadrupolar and electric dipole (or electric polarization) operators
have the same symmetry properties. Thus, a new channel opens to
excite these modes as the electric field of the incident light can
directly couple to the spin quadrupolar operators, response
expressed by the $\text{Im}\chi^{ee}_{\alpha\alpha}(\omega)$
\cite{Miyahara2011}. Indeed, $d_-$ and $c_+$ modes with only
magnetic quadrupol moment are excited this way and remain silent in
$\text{Im}\chi^{mm}_{\alpha\alpha}(\omega)$ irrespective of the
$\Lambda/J$ ratio. The dynamical electric susceptibility
$\text{Im}\chi^{ee}_{\alpha\alpha}(\omega)$ shows a strong response
 for most of the modes, as shown in Fig.~\ref{fig:modes_omega}(c).

Our model describes well the magnetic field dependence of the
spin-wave frequencies in Ba$_2$CoGe$_2$O$_7$, see
Fig.~\ref{fig:spectra}. From the fit of the experimental data we
obtain $\Lambda=13.4\rm\,K$,
$J=2.3\rm\,K$, $J_z=1.8\rm\,K$, $g_{zz}=2.1$ and
$g_{xx}=g_{yy}=2.3$. Magnetic field larger than $16\rm\,T$ in the easy plane is
strong enough to drive a transition from a canted AF to an almost
saturated magnet. This is observed as a kink in the 1\,THz modes at
$B_{\rm dc}\approx16\rm\,T$ [Fig.~\ref{fig:spectra}(b)].
The theory also predicts the onset of fully saturated phase
for $B_{\rm dc}>36\rm\,T$ applied perpendicular to the easy plane (along the tetragonal axis),
inducing a gap in the Goldstone mode [Fig.~\ref{fig:spectra}(a)], in agreement with Ref.~\onlinecite{NHMFL2010}.
The V-shape splitting of the 1\,THz mode and the avoided crossing at
$B_{\rm dc}\approx12$\,T for fields parallel to the tetragonal axis
is also reproduced correctly [see Fig.~\ref{fig:spectra}(a) and
(c)]. The lowest-lying mode of the $f\sim1$\,THz branch
is theoretically predicted to be weak [see dotted grey line in
Fig.~\ref{fig:modes_omega}(b)] and does not appear in the experimental
spectra. The only feature not explained by the model is the
splitting of the $f\sim0.5$\,THz resonance above $B_{\rm dc}=5$\,T
for fields perpendicular to the [001] axis.

The analytical solution of a pure spin Hamiltonian
[see Eq.~(\ref{eq:Hamiltonian_1})] enabled us to fully characterize the
excited states in terms of spin and polarization dynamics, implying that
the electric polarization adiabatically follows the sublattice
magnetization vector and does not have its ``own'' dynamics in the
energy range of interest.
The motion of the sublattice magnetization and the local polarization in
zero field is visualized in Fig.~\ref{fig:modes_anim} for the Goldstone
mode and for a spin-stretching mode. The Goldstone mode is associated
with the oscillation of the polarization along the tetragonal axis and
has a direct connection with the dc magnetoelectric effect. The
spin-stretching mode shows more complex polarization dynamics
\footnote{See supplementary material at [URL] for an animated
representation of each mode.}, e.g., for $\Lambda \to 0$ the polarization
still oscillates even though the magnetic moment is frozen.

Our theory describes the unconventional spin-wave
excitations in Ba$_2$CoGe$_2$O$_7$
and provides a guide for spin-wave spectroscopy in a broad class of ordered magnets
with strong magnetic anisotropy and/or a non-centrosymmetric lattice structure. 
We expect unconventional spin excitations
to emerge in the dynamic magnetic
susceptibility whenever a large single-ion anisotropy is present in
a $S>1/2$ system. 
Moreover, if the inversion symmetry of the crystal is
broken, these new modes can have a dielectric response even in the
absence of magnetic anisotropy.
They should be detected by THz
light absorption or inelastic neutron scattering via the induced
magnetic and/or electric dipole moment.

\begin{acknowledgments}
We are grateful for stimulating discussions with S.~Miyahara,
N.~Furukawa, N.~Kida, Y.~Onose, and B.~N\'afr\'adi. 
This work was supported by Hungarian OTKA under Grant Nos. PD75615, CNK80991, K73361,
K73455, and NN76727, Bolyai program, T\'AMOP-4.2.1/B-09/1/KMR-2010-0002,
by EuroMagNET II under the EU contract number 228043, by Estonian
Ministry of Education and Research Grant SF0690029s09, Estonian
Science Foundation Grants ETF8170 and ETF8703, and the bilateral
programme of the Estonian and Hungarian Academies of Science.
\end{acknowledgments}

\bibliography{flavor_waves_v4_5}

\begin{thebibliography}{27}%
\makeatletter
\providecommand \@ifxundefined [1]{%
 \@ifx{#1\undefined}
}%
\providecommand \@ifnum [1]{%
 \ifnum #1\expandafter \@firstoftwo
 \else \expandafter \@secondoftwo
 \fi
}%
\providecommand \@ifx [1]{%
 \ifx #1\expandafter \@firstoftwo
 \else \expandafter \@secondoftwo
 \fi
}%
\providecommand \natexlab [1]{#1}%
\providecommand \enquote  [1]{``#1''}%
\providecommand \bibnamefont  [1]{#1}%
\providecommand \bibfnamefont [1]{#1}%
\providecommand \citenamefont [1]{#1}%
\providecommand \href@noop [0]{\@secondoftwo}%
\providecommand \href [0]{\begingroup \@sanitize@url \@href}%
\providecommand \@href[1]{\@@startlink{#1}\@@href}%
\providecommand \@@href[1]{\endgroup#1\@@endlink}%
\providecommand \@sanitize@url [0]{\catcode `\\12\catcode `\$12\catcode
  `\&12\catcode `\#12\catcode `\^12\catcode `\_12\catcode `\%12\relax}%
\providecommand \@@startlink[1]{}%
\providecommand \@@endlink[0]{}%
\providecommand \url  [0]{\begingroup\@sanitize@url \@url }%
\providecommand \@url [1]{\endgroup\@href {#1}{\urlprefix }}%
\providecommand \urlprefix  [0]{URL }%
\providecommand \Eprint [0]{\href }%
\providecommand \doibase [0]{http://dx.doi.org/}%
\providecommand \selectlanguage [0]{\@gobble}%
\providecommand \bibinfo  [0]{\@secondoftwo}%
\providecommand \bibfield  [0]{\@secondoftwo}%
\providecommand \translation [1]{[#1]}%
\providecommand \BibitemOpen [0]{}%
\providecommand \bibitemStop [0]{}%
\providecommand \bibitemNoStop [0]{.\EOS\space}%
\providecommand \EOS [0]{\spacefactor3000\relax}%
\providecommand \BibitemShut  [1]{\csname bibitem#1\endcsname}%
\let\auto@bib@innerbib\@empty
%</preamble>
\bibitem [{\citenamefont {Van~Kranendonk}\ and\ \citenamefont
  {Van~Vleck}(1958)}]{VanKranendonk1958}%
  \BibitemOpen
  \bibfield  {author} {\bibinfo {author} {\bibfnamefont {J.}~\bibnamefont
  {Van~Kranendonk}}\ and\ \bibinfo {author} {\bibfnamefont {J.~H.}\
  \bibnamefont {Van~Vleck}},\ }\href {\doibase 10.1103/RevModPhys.30.1}
  {\bibfield  {journal} {\bibinfo  {journal} {Rev. Mod. Phys.}\ }\textbf
  {\bibinfo {volume} {30}},\ \bibinfo {pages} {1} (\bibinfo {year}
  {1958})}\BibitemShut {NoStop}%
\bibitem [{\citenamefont {Turov}(1965)}]{Turov}%
  \BibitemOpen
  \bibfield  {author} {\bibinfo {author} {\bibfnamefont {E.}~\bibnamefont
  {Turov}},\ }\href@noop {} {\emph {\bibinfo {title} {Physical Properties of
  Magnetically Ordered Crystals}}}\ (\bibinfo  {publisher} {Academic Press},\
  \bibinfo {address} {New York},\ \bibinfo {year} {1965})\BibitemShut {NoStop}%
\bibitem [{\citenamefont {Shiina}\ \emph {et~al.}(2003)\citenamefont {Shiina},
  \citenamefont {Shiba}, \citenamefont {Thalmeier}, \citenamefont {Takahashi},\
  and\ \citenamefont {Sakai}}]{Shiina2003}%
  \BibitemOpen
  \bibfield  {author} {\bibinfo {author} {\bibfnamefont {R.}~\bibnamefont
  {Shiina}}, \bibinfo {author} {\bibfnamefont {H.}~\bibnamefont {Shiba}},
  \bibinfo {author} {\bibfnamefont {P.}~\bibnamefont {Thalmeier}}, \bibinfo
  {author} {\bibfnamefont {A.}~\bibnamefont {Takahashi}}, \ and\ \bibinfo
  {author} {\bibfnamefont {O.}~\bibnamefont {Sakai}},\ }\href {\doibase
  10.1143/JPSJ.72.1216} {\bibfield  {journal} {\bibinfo  {journal} {Journal of
  the Physical Society of Japan}\ }\textbf {\bibinfo {volume} {72}},\ \bibinfo
  {pages} {1216} (\bibinfo {year} {2003})}\BibitemShut {NoStop}%
\bibitem [{\citenamefont {Carretta}\ \emph {et~al.}(2010)\citenamefont
  {Carretta}, \citenamefont {Santini}, \citenamefont {Caciuffo},\ and\
  \citenamefont {Amoretti}}]{Carretta2010}%
  \BibitemOpen
  \bibfield  {author} {\bibinfo {author} {\bibfnamefont {S.}~\bibnamefont
  {Carretta}}, \bibinfo {author} {\bibfnamefont {P.}~\bibnamefont {Santini}},
  \bibinfo {author} {\bibfnamefont {R.}~\bibnamefont {Caciuffo}}, \ and\
  \bibinfo {author} {\bibfnamefont {G.}~\bibnamefont {Amoretti}},\ }\href
  {\doibase 10.1103/PhysRevLett.105.167201} {\bibfield  {journal} {\bibinfo
  {journal} {Phys. Rev. Lett.}\ }\textbf {\bibinfo {volume} {105}},\ \bibinfo
  {pages} {167201} (\bibinfo {year} {2010})}\BibitemShut {NoStop}%
\bibitem [{\citenamefont {K\'ezsm\'arki}\ \emph {et~al.}(2011)\citenamefont
  {K\'ezsm\'arki}, \citenamefont {Kida}, \citenamefont {Murakawa},
  \citenamefont {Bord\'acs}, \citenamefont {Onose},\ and\ \citenamefont
  {Tokura}}]{Kezsmarki2011}%
  \BibitemOpen
  \bibfield  {author} {\bibinfo {author} {\bibfnamefont {I.}~\bibnamefont
  {K\'ezsm\'arki}}, \bibinfo {author} {\bibfnamefont {N.}~\bibnamefont {Kida}},
  \bibinfo {author} {\bibfnamefont {H.}~\bibnamefont {Murakawa}}, \bibinfo
  {author} {\bibfnamefont {S.}~\bibnamefont {Bord\'acs}}, \bibinfo {author}
  {\bibfnamefont {Y.}~\bibnamefont {Onose}}, \ and\ \bibinfo {author}
  {\bibfnamefont {Y.}~\bibnamefont {Tokura}},\ }\href {\doibase
  10.1103/PhysRevLett.106.057403} {\bibfield  {journal} {\bibinfo  {journal}
  {Phys. Rev. Lett.}\ }\textbf {\bibinfo {volume} {106}},\ \bibinfo {pages}
  {057403} (\bibinfo {year} {2011})}\BibitemShut {NoStop}%
\bibitem [{Note1()}]{Note1}%
  \BibitemOpen
  \bibinfo {note} {C. de la Cruz, private communication}\BibitemShut {NoStop}%
\bibitem [{\citenamefont {Zheludev}\ \emph {et~al.}(2003)\citenamefont
  {Zheludev}, \citenamefont {Sato}, \citenamefont {Masuda}, \citenamefont
  {Uchinokura}, \citenamefont {Shirane},\ and\ \citenamefont
  {Roessli}}]{Zheludev2003}%
  \BibitemOpen
  \bibfield  {author} {\bibinfo {author} {\bibfnamefont {A.}~\bibnamefont
  {Zheludev}}, \bibinfo {author} {\bibfnamefont {T.}~\bibnamefont {Sato}},
  \bibinfo {author} {\bibfnamefont {T.}~\bibnamefont {Masuda}}, \bibinfo
  {author} {\bibfnamefont {K.}~\bibnamefont {Uchinokura}}, \bibinfo {author}
  {\bibfnamefont {G.}~\bibnamefont {Shirane}}, \ and\ \bibinfo {author}
  {\bibfnamefont {B.}~\bibnamefont {Roessli}},\ }\href {\doibase
  10.1103/PhysRevB.68.024428} {\bibfield  {journal} {\bibinfo  {journal} {Phys.
  Rev. B}\ }\textbf {\bibinfo {volume} {68}},\ \bibinfo {pages} {024428}
  (\bibinfo {year} {2003})}\BibitemShut {NoStop}%
\bibitem [{\citenamefont {Murakawa}\ \emph {et~al.}(2010)\citenamefont
  {Murakawa}, \citenamefont {Onose}, \citenamefont {Miyahara}, \citenamefont
  {Furukawa},\ and\ \citenamefont {Tokura}}]{Murakawa2010}%
  \BibitemOpen
  \bibfield  {author} {\bibinfo {author} {\bibfnamefont {H.}~\bibnamefont
  {Murakawa}}, \bibinfo {author} {\bibfnamefont {Y.}~\bibnamefont {Onose}},
  \bibinfo {author} {\bibfnamefont {S.}~\bibnamefont {Miyahara}}, \bibinfo
  {author} {\bibfnamefont {N.}~\bibnamefont {Furukawa}}, \ and\ \bibinfo
  {author} {\bibfnamefont {Y.}~\bibnamefont {Tokura}},\ }\href {\doibase
  10.1103/PhysRevLett.105.137202} {\bibfield  {journal} {\bibinfo  {journal}
  {Phys. Rev. Lett.}\ }\textbf {\bibinfo {volume} {105}},\ \bibinfo {pages}
  {137202} (\bibinfo {year} {2010})}\BibitemShut {NoStop}%
\bibitem [{\citenamefont {Yi}\ \emph {et~al.}(2008)\citenamefont {Yi},
  \citenamefont {Choi}, \citenamefont {Lee},\ and\ \citenamefont
  {Cheong}}]{Yi2008}%
  \BibitemOpen
  \bibfield  {author} {\bibinfo {author} {\bibfnamefont {H.~T.}\ \bibnamefont
  {Yi}}, \bibinfo {author} {\bibfnamefont {Y.~J.}\ \bibnamefont {Choi}},
  \bibinfo {author} {\bibfnamefont {S.}~\bibnamefont {Lee}}, \ and\ \bibinfo
  {author} {\bibfnamefont {S.-W.}\ \bibnamefont {Cheong}},\ }\href {\doibase
  10.1063/1.2937110} {\bibfield  {journal} {\bibinfo  {journal} {Appl. Phys.
  Lett.}\ }\textbf {\bibinfo {volume} {92}},\ \bibinfo {eid} {212904} (\bibinfo
  {year} {2008})}\BibitemShut {NoStop}%
\bibitem [{\citenamefont {{Bordacs}}\ \emph {et~al.}(2011)\citenamefont
  {{Bordacs}}, \citenamefont {{Kezsmarki}}, \citenamefont {{Szaller}},
  \citenamefont {{Demko}}, \citenamefont {{Kida}}, \citenamefont {{Murakawa}},
  \citenamefont {{Onose}}, \citenamefont {{Shimano}}, \citenamefont {{Room}},
  \citenamefont {{Nagel}}, \citenamefont {{Miyahara}}, \citenamefont
  {{Furukawa}},\ and\ \citenamefont {{Tokura}}}]{Bordacs}%
  \BibitemOpen
  \bibfield  {author} {\bibinfo {author} {\bibfnamefont {S.}~\bibnamefont
  {{Bordacs}}}, \bibinfo {author} {\bibfnamefont {I.}~\bibnamefont
  {{Kezsmarki}}}, \bibinfo {author} {\bibfnamefont {D.}~\bibnamefont
  {{Szaller}}}, \bibinfo {author} {\bibfnamefont {L.}~\bibnamefont {{Demko}}},
  \bibinfo {author} {\bibfnamefont {N.}~\bibnamefont {{Kida}}}, \bibinfo
  {author} {\bibfnamefont {H.}~\bibnamefont {{Murakawa}}}, \bibinfo {author}
  {\bibfnamefont {Y.}~\bibnamefont {{Onose}}}, \bibinfo {author} {\bibfnamefont
  {R.}~\bibnamefont {{Shimano}}}, \bibinfo {author} {\bibfnamefont
  {T.}~\bibnamefont {{Room}}}, \bibinfo {author} {\bibfnamefont
  {U.}~\bibnamefont {{Nagel}}}, \bibinfo {author} {\bibfnamefont
  {S.}~\bibnamefont {{Miyahara}}}, \bibinfo {author} {\bibfnamefont
  {N.}~\bibnamefont {{Furukawa}}}, \ and\ \bibinfo {author} {\bibfnamefont
  {Y.}~\bibnamefont {{Tokura}}},\ }\href@noop {} {\bibfield  {journal}
  {\bibinfo  {journal} {ArXiv e-prints}\ } (\bibinfo {year} {2011})},\ \Eprint
  {http://arxiv.org/abs/1109.1597} {arXiv:1109.1597 [cond-mat.str-el]}
  \BibitemShut {NoStop}%
\bibitem [{\citenamefont {Miyahara}\ and\ \citenamefont
  {Furukawa}(2011)}]{Miyahara2011}%
  \BibitemOpen
  \bibfield  {author} {\bibinfo {author} {\bibfnamefont {S.}~\bibnamefont
  {Miyahara}}\ and\ \bibinfo {author} {\bibfnamefont {N.}~\bibnamefont
  {Furukawa}},\ }\href {\doibase 10.1143/JPSJ.80.073708} {\bibfield  {journal}
  {\bibinfo  {journal} {Journal of the Physical Society of Japan}\ }\textbf
  {\bibinfo {volume} {80}},\ \bibinfo {pages} {073708} (\bibinfo {year}
  {2011})}\BibitemShut {NoStop}%
\bibitem [{\citenamefont {Toledano}\ \emph {et~al.}(2011)\citenamefont
  {Toledano}, \citenamefont {Khalyavin},\ and\ \citenamefont
  {Chapon}}]{Toledano2011}%
  \BibitemOpen
  \bibfield  {author} {\bibinfo {author} {\bibfnamefont {P.}~\bibnamefont
  {Toledano}}, \bibinfo {author} {\bibfnamefont {D.~D.}\ \bibnamefont
  {Khalyavin}}, \ and\ \bibinfo {author} {\bibfnamefont {L.~C.}\ \bibnamefont
  {Chapon}},\ }\href {\doibase 10.1103/PhysRevB.84.094421} {\bibfield
  {journal} {\bibinfo  {journal} {Phys. Rev. B}\ }\textbf {\bibinfo {volume}
  {84}},\ \bibinfo {pages} {094421} (\bibinfo {year} {2011})}\BibitemShut
  {NoStop}%
\bibitem [{\citenamefont {Yamauchi}\ \emph {et~al.}(2011)\citenamefont
  {Yamauchi}, \citenamefont {Barone},\ and\ \citenamefont
  {Picozzi}}]{Yamauchi2011}%
  \BibitemOpen
  \bibfield  {author} {\bibinfo {author} {\bibfnamefont {K.}~\bibnamefont
  {Yamauchi}}, \bibinfo {author} {\bibfnamefont {P.}~\bibnamefont {Barone}}, \
  and\ \bibinfo {author} {\bibfnamefont {S.}~\bibnamefont {Picozzi}},\ }\href
  {\doibase 10.1103/PhysRevB.84.165137} {\bibfield  {journal} {\bibinfo
  {journal} {Phys. Rev. B}\ }\textbf {\bibinfo {volume} {84}},\ \bibinfo
  {pages} {165137} (\bibinfo {year} {2011})}\BibitemShut {NoStop}%
\bibitem [{\citenamefont {Akaki}\ \emph {et~al.}(2009)\citenamefont {Akaki},
  \citenamefont {Tozawa}, \citenamefont {Akahoshi},\ and\ \citenamefont
  {Kuwahara}}]{Akai2009}%
  \BibitemOpen
  \bibfield  {author} {\bibinfo {author} {\bibfnamefont {M.}~\bibnamefont
  {Akaki}}, \bibinfo {author} {\bibfnamefont {J.}~\bibnamefont {Tozawa}},
  \bibinfo {author} {\bibfnamefont {D.}~\bibnamefont {Akahoshi}}, \ and\
  \bibinfo {author} {\bibfnamefont {H.}~\bibnamefont {Kuwahara}},\ }\href
  {\doibase 10.1063/1.3147195} {\bibfield  {journal} {\bibinfo  {journal}
  {Applied Physics Letters}\ }\textbf {\bibinfo {volume} {94}},\ \bibinfo {eid}
  {212904} (\bibinfo {year} {2009})}\BibitemShut {NoStop}%
\bibitem [{\citenamefont {Akaki}\ \emph {et~al.}(2010)\citenamefont {Akaki},
  \citenamefont {Tozawa}, \citenamefont {Hitomi}, \citenamefont {Akahoshi},\
  and\ \citenamefont {Kuwahara}}]{Akai2010}%
  \BibitemOpen
  \bibfield  {author} {\bibinfo {author} {\bibfnamefont {M.}~\bibnamefont
  {Akaki}}, \bibinfo {author} {\bibfnamefont {J.}~\bibnamefont {Tozawa}},
  \bibinfo {author} {\bibfnamefont {M.}~\bibnamefont {Hitomi}}, \bibinfo
  {author} {\bibfnamefont {D.}~\bibnamefont {Akahoshi}}, \ and\ \bibinfo
  {author} {\bibfnamefont {H.}~\bibnamefont {Kuwahara}},\ }\href
  {http://stacks.iop.org/1742-6596/200/i=1/a=012003} {\bibfield  {journal}
  {\bibinfo  {journal} {Journal of Physics: Conference Series}\ }\textbf
  {\bibinfo {volume} {200}},\ \bibinfo {pages} {012003} (\bibinfo {year}
  {2010})}\BibitemShut {NoStop}%
\bibitem [{Note2()}]{Note2}%
  \BibitemOpen
  \bibinfo {note} {H. Murakawa, private communication}\BibitemShut {NoStop}%
\bibitem [{\citenamefont {Romh\'anyi}\ \emph
  {et~al.}(2011{\natexlab{a}})\citenamefont {Romh\'anyi}, \citenamefont
  {Pollmann},\ and\ \citenamefont {Penc}}]{PhysRevB.84.184427}%
  \BibitemOpen
  \bibfield  {author} {\bibinfo {author} {\bibfnamefont {J.}~\bibnamefont
  {Romh\'anyi}}, \bibinfo {author} {\bibfnamefont {F.}~\bibnamefont
  {Pollmann}}, \ and\ \bibinfo {author} {\bibfnamefont {K.}~\bibnamefont
  {Penc}},\ }\href {\doibase 10.1103/PhysRevB.84.184427} {\bibfield  {journal}
  {\bibinfo  {journal} {Phys. Rev. B}\ }\textbf {\bibinfo {volume} {84}},\
  \bibinfo {pages} {184427} (\bibinfo {year} {2011}{\natexlab{a}})}\BibitemShut
  {NoStop}%
\bibitem [{\citenamefont {Romh\'anyi}\ \emph
  {et~al.}(2011{\natexlab{b}})\citenamefont {Romh\'anyi}, \citenamefont
  {Lajk\'o},\ and\ \citenamefont {Penc}}]{PhysRevB.84.224419}%
  \BibitemOpen
  \bibfield  {author} {\bibinfo {author} {\bibfnamefont {J.}~\bibnamefont
  {Romh\'anyi}}, \bibinfo {author} {\bibfnamefont {M.}~\bibnamefont {Lajk\'o}},
  \ and\ \bibinfo {author} {\bibfnamefont {K.}~\bibnamefont {Penc}},\ }\href
  {\doibase 10.1103/PhysRevB.84.224419} {\bibfield  {journal} {\bibinfo
  {journal} {Phys. Rev. B}\ }\textbf {\bibinfo {volume} {84}},\ \bibinfo
  {pages} {224419} (\bibinfo {year} {2011}{\natexlab{b}})}\BibitemShut
  {NoStop}%
\bibitem [{Note3()}]{Note3}%
  \BibitemOpen
  \bibinfo {note} {See supplementary material at [URL] for the exact form of
  the transformed Hamiltonian and the details of the calculation.}\BibitemShut
  {Stop}%
\bibitem [{\citenamefont {Joshi}\ \emph {et~al.}(1999)\citenamefont {Joshi},
  \citenamefont {Ma}, \citenamefont {Mila}, \citenamefont {Shi},\ and\
  \citenamefont {Zhang}}]{PhysRevB.60.6584}%
  \BibitemOpen
  \bibfield  {author} {\bibinfo {author} {\bibfnamefont {A.}~\bibnamefont
  {Joshi}}, \bibinfo {author} {\bibfnamefont {M.}~\bibnamefont {Ma}}, \bibinfo
  {author} {\bibfnamefont {F.}~\bibnamefont {Mila}}, \bibinfo {author}
  {\bibfnamefont {D.~N.}\ \bibnamefont {Shi}}, \ and\ \bibinfo {author}
  {\bibfnamefont {F.~C.}\ \bibnamefont {Zhang}},\ }\href {\doibase
  10.1103/PhysRevB.60.6584} {\bibfield  {journal} {\bibinfo  {journal} {Phys.
  Rev. B}\ }\textbf {\bibinfo {volume} {60}},\ \bibinfo {pages} {6584}
  (\bibinfo {year} {1999})}\BibitemShut {NoStop}%
\bibitem [{\citenamefont {Matsumoto}\ \emph {et~al.}(2004)\citenamefont
  {Matsumoto}, \citenamefont {Normand}, \citenamefont {Rice},\ and\
  \citenamefont {Sigrist}}]{PhysRevB.69.054423}%
  \BibitemOpen
  \bibfield  {author} {\bibinfo {author} {\bibfnamefont {M.}~\bibnamefont
  {Matsumoto}}, \bibinfo {author} {\bibfnamefont {B.}~\bibnamefont {Normand}},
  \bibinfo {author} {\bibfnamefont {T.~M.}\ \bibnamefont {Rice}}, \ and\
  \bibinfo {author} {\bibfnamefont {M.}~\bibnamefont {Sigrist}},\ }\href
  {\doibase 10.1103/PhysRevB.69.054423} {\bibfield  {journal} {\bibinfo
  {journal} {Phys. Rev. B}\ }\textbf {\bibinfo {volume} {69}},\ \bibinfo
  {pages} {054423} (\bibinfo {year} {2004})}\BibitemShut {NoStop}%
\bibitem [{\citenamefont {Papanicolaou}(1984)}]{N1984281}%
  \BibitemOpen
  \bibfield  {author} {\bibinfo {author} {\bibfnamefont {N.}~\bibnamefont
  {Papanicolaou}},\ }\href {\doibase 10.1016/0550-3213(84)90268-2} {\bibfield
  {journal} {\bibinfo  {journal} {Nuclear Physics B}\ }\textbf {\bibinfo
  {volume} {240}},\ \bibinfo {pages} {281 } (\bibinfo {year}
  {1984})}\BibitemShut {NoStop}%
\bibitem [{\citenamefont {Papanicolaou}(1988)}]{N1988367}%
  \BibitemOpen
  \bibfield  {author} {\bibinfo {author} {\bibfnamefont {N.}~\bibnamefont
  {Papanicolaou}},\ }\href {\doibase 10.1016/0550-3213(88)90073-9} {\bibfield
  {journal} {\bibinfo  {journal} {Nuclear Physics B}\ }\textbf {\bibinfo
  {volume} {305}},\ \bibinfo {pages} {367 } (\bibinfo {year}
  {1988})}\BibitemShut {NoStop}%
\bibitem [{\citenamefont {Onufrieva}(1985)}]{ISI:A1985AXC5300038}%
  \BibitemOpen
  \bibfield  {author} {\bibinfo {author} {\bibfnamefont {F.}~\bibnamefont
  {Onufrieva}},\ }\href@noop {} {\bibfield  {journal} {\bibinfo  {journal}
  {Zhurnal Eksperimentalnoi i Teoreticheskoi Fiziki}\ }\textbf {\bibinfo
  {volume} {89}},\ \bibinfo {pages} {2270} (\bibinfo {year}
  {1985})}\BibitemShut {NoStop}%
\bibitem [{\citenamefont {Chubukov}(1990)}]{0953-8984-2-6-018}%
  \BibitemOpen
  \bibfield  {author} {\bibinfo {author} {\bibfnamefont {A.~V.}\ \bibnamefont
  {Chubukov}},\ }\href {http://stacks.iop.org/0953-8984/2/i=6/a=018} {\bibfield
   {journal} {\bibinfo  {journal} {Journal of Physics: Condensed Matter}\
  }\textbf {\bibinfo {volume} {2}},\ \bibinfo {pages} {1593} (\bibinfo {year}
  {1990})}\BibitemShut {NoStop}%
\bibitem [{\citenamefont {Kim}\ \emph {et~al.}(2010)\citenamefont {Kim},
  \citenamefont {Khim}, \citenamefont {Chun}, \citenamefont {Kim},
  \citenamefont {Choi}, \citenamefont {Jo}, \citenamefont {Balicas},
  \citenamefont {Harrison}, \citenamefont {Yi}, \citenamefont {Cheong},
  \citenamefont {Han},\ and\ \citenamefont {Batista}}]{NHMFL2010}%
  \BibitemOpen
  \bibfield  {author} {\bibinfo {author} {\bibfnamefont {J.~W.}\ \bibnamefont
  {Kim}}, \bibinfo {author} {\bibfnamefont {S.~H.}\ \bibnamefont {Khim}},
  \bibinfo {author} {\bibfnamefont {S.~H.}\ \bibnamefont {Chun}}, \bibinfo
  {author} {\bibfnamefont {K.~H.}\ \bibnamefont {Kim}}, \bibinfo {author}
  {\bibfnamefont {E.}~\bibnamefont {Choi}}, \bibinfo {author} {\bibfnamefont
  {Y.}~\bibnamefont {Jo}}, \bibinfo {author} {\bibfnamefont {L.}~\bibnamefont
  {Balicas}}, \bibinfo {author} {\bibfnamefont {N.}~\bibnamefont {Harrison}},
  \bibinfo {author} {\bibfnamefont {H.}~\bibnamefont {Yi}}, \bibinfo {author}
  {\bibfnamefont {S.-W.}\ \bibnamefont {Cheong}}, \bibinfo {author}
  {\bibfnamefont {J.~H.}\ \bibnamefont {Han}}, \ and\ \bibinfo {author}
  {\bibfnamefont {C.~D.}\ \bibnamefont {Batista}},\ }\href@noop {} {\emph
  {\bibinfo {title} {Investigation of a Quantum Critical Point in Multiferroic
  Ba$_2$CoGe$_2$O$_7$}}},\ \bibinfo {type} {Annual Report}\ \bibinfo {number}
  {p. 20}\ (\bibinfo  {institution} {National High Magnetic Field Laboratory},\
  \bibinfo {year} {2010})\BibitemShut {NoStop}%
\bibitem [{Note4()}]{Note4}%
  \BibitemOpen
  \bibinfo {note} {See supplementary material at [URL] for an animated
  representation of each mode.}\BibitemShut {Stop}%
\end{thebibliography}%

\clearpage
\section{Supplement}

%\date{\today}

\maketitle

In this supplement we show in more details the multiboson spin-wave theory.
For simplicity, we consider the case when the magnetic field is in the easy plane. For magnetic field perpendicular to the easy-plane the expressions become rather complicated, and the energy of the modes that was shown in Fig.~1 can be obtained numerically.

\subsection{Bosonic representation of the spin operators}

First we introduce the bosons $\alpha^\dagger_m$ that create the  $S^z=m$ states of the $S=3/2$ spin, i.e. $|m \rangle = \alpha^\dagger_{m} |\text{vacuum}\rangle$. The number of bosons on each site is conserved,
$\sum_{m} \alpha^{\dagger}_{m} \alpha^{\phantom{\dagger}}_{m} = M$, and $M=1$ for the $S=3/2$ spin.
Using the four $\alpha$ bosons, the spin operators can all be expressed as quadratic forms, for example
\begin{eqnarray}
S^z &=& \sum_{m=-3/2}^{3/2} m \alpha^{\dagger}_{m} \alpha^{\phantom{\dagger}}_{m},\\
(S^z)^2 &=& \sum_{m=-3/2}^{3/2} m^2 \alpha^{\dagger}_{m} \alpha^{\phantom{\dagger}}_{m},\\
S^+ &=& \sqrt{3} \left(
  \alpha^{\dagger}_{3/2} \alpha^{\phantom{\dagger}}_{1/2} \!+\!
    \alpha^{\dagger}_{-1/2} \alpha^{\phantom{\dagger}}_{-3/2} \right) 
%    \nonumber\\&& 
 + 2 \alpha^{\dagger}_{1/2} \alpha^{\phantom{\dagger}}_{-1/2}, \\
(S^+)^2 &=& 2 \sqrt{3} \left(
  \alpha^{\dagger}_{3/2} \alpha^{\phantom{\dagger}}_{-1/2} +
    \alpha^{\dagger}_{1/2} \alpha^{\phantom{\dagger}}_{-3/2} \right).
\end{eqnarray}

Next, we apply an SU(4) rotation in the space of $\alpha^\dagger_m$ bosons:
\begin{widetext}
\begin{subequations}
\begin{eqnarray}
a^{\dagger}_A = a^{\dagger}_{0,A} &=& \frac{1}{\sqrt{6 \eta^2+2}} \left[
e^{\frac{3}{2} i \varphi_A } \alpha^{\dagger}_{-3/2}
+e^{-\frac{3}{2} i \varphi_A} \alpha^{\dagger}_{3/2}
+\sqrt{3} \eta \left( e^{\frac{1}{2} i \varphi_A}  \alpha^{\dagger}_{-1/2}
+ e^{-\frac{1}{2}i \varphi_A }  \alpha^{\dagger}_{1/2} \right)
\right],
\label{eq:aboson}\\
b^{\dagger}_A  = a^{\dagger}_{1,A} &=& \frac{1}{\sqrt{14 \eta^2-8 \eta +2}} \left[
\sqrt{3} \eta \left( e^{\frac{3}{2} i \varphi_A } \alpha^{\dagger}_{-3/2}
  - e^{-\frac{3}{2} i \varphi_A } \alpha^{\dagger}_{3/2} \right)
+(2 \eta -1) \left( 
  e^{\frac{1}{2} i \varphi_A } \alpha^{\dagger}_{-1/2}
  -e^{-\frac{1}{2} i \varphi_A } \alpha^{\dagger}_{1/2} 
  \right)
\right],
\label{eq:bboson}\\
c^{\dagger}_A  = a^{\dagger}_{2,A}  &=& 
\frac{1}{\sqrt{6 \eta^2+2}}
\left[
\sqrt{3} \eta \left(
  e^{\frac{3}{2} i \varphi_A } \alpha^{\dagger}_{-3/2} 
+ e^{-\frac{3}{2} i \varphi_A } \alpha^{\dagger}_{3/2}
\right)
- \left( e^{\frac{1}{2} i \varphi_A }\alpha^{\dagger}_{-1/2}
+ e^{-\frac{1}{2} i \varphi_A }\alpha^{\dagger}_{1/2} \right)
\right],
\label{eq:cboson}\\
d^{\dagger}_A  = a^{\dagger}_{3,A}  &=& \frac{1}{\sqrt{14 \eta^2-8 \eta +2}} \left[
(2 \eta -1) \left( e^{\frac{3}{2} i \varphi_A } \alpha^{\dagger}_{-3/2}
-e^{-\frac{3}{2} i \varphi_A } \alpha^{\dagger}_{3/2}  \right)
-\sqrt{3} \eta \left(e^{\frac{1}{2} i \varphi_A } \alpha^{\dagger}_{-1/2}
 - e^{-\frac{1}{2} i \varphi_A }  \alpha^{\dagger}_{1/2} \right)
\right],
\label{eq:dboson}
\end{eqnarray}
\end{subequations}
\end{widetext}
and analogous expressions hold for the bosons on the $B$-sites.

\subsection{Variational solution}

In the rotated basis the variational wave functionn corresponds to the 
$a^\dagger_A|\mbox{vacuum}\rangle$. 
The energy per site, as a function of $\eta$ and $\varphi=\varphi_A=-\varphi_B$, reads
\begin{eqnarray}
\frac{E(\eta,\varphi)}{N} &=& \frac{3}{4} \frac{ \left(\eta^2+3\right)}{\left(3 \eta^2+1\right)} \Lambda
+\frac{18 \eta^2 (\eta +1)^2 }{\left(3 \eta^2+1\right)^2} J \cos 2 \varphi
\nonumber\\
&&-\frac{3 \eta  (\eta +1)}{3 \eta^2+1}  g_{xx} h_x \cos \varphi .
\label{eq:SF_en0}
 \end{eqnarray}
 Minimizing the $E(\eta,\varphi)$ with respect to variational parameters $\eta$ and $\varphi$, we get 
two solutions: 
(i) canted N\'eel-state, defined via the following set of equations:
\begin{subequations}
\begin{eqnarray}
\Lambda &=& \frac{3 (3 \eta +1) \left(\eta^2-1\right)}{3 \eta^2+1} J,
\\
g_{xx} h_x &=& \frac{24 \eta  (\eta+1)
   }{\left(3 \eta^2+1\right) } J \cos \varphi .
   \label{eq:eta_sol}
\end{eqnarray}
\label{eq:eta_sol}
\end{subequations}
The spins cant in the direction of the field keeping the $\eta$ parameter unchanged. The limiting cases for $\eta$ are
\begin{equation}
 \eta = \left\{
 \begin{array}{lc}
 {\displaystyle 1+\frac{\Lambda }{6 J} +O\left(\Lambda^2/J^2\right)}, \; &\mbox{if $\Lambda \ll J$}; \\
 \\
{\displaystyle \frac{\Lambda }{3 J}-\frac{1}{3} +O\left(J/\Lambda\right)}, \; &\mbox{if $\Lambda \gg J$} .
\end{array}
 \right.
\end{equation}
(ii) For high enough magnetic field, the spins in the A and B sublattice become equal and parallel to the field, setting $\varphi = 0$, and $\eta$ is obtained from:
\begin{equation}
\Lambda = \frac{(\eta-1) (3\eta+1)}{4\eta} g_{xx} h_x 
- \frac{3 (3 \eta +1) \left(\eta^2-1\right)}{3 \eta^2+1} J.
 \label{eq:eta_sol_sat}
\end{equation}
$\eta \to 1$ as the field $h_x \to \infty$.

While the usual procedure is to solve the equations for $\eta$ and $\varphi$, we prefer to express the $\Lambda$ as a function of $\eta$ in the following.   
  
\pagebreak
\subsection{Multiboson spin-waves}

The $b_X$, $c_X$, and $d_X$ (i.e. $a_{\nu,X}$, with $X = A,B$) bosons in Eqs.~(\ref{eq:bboson})--(\ref{eq:dboson}) take the role of the Holstein-Primakoff bosons. After the $1/M$ expansion, the spin dipole operators are
\begin{widetext}
\begin{subequations}
\begin{eqnarray}
S_X^x &=& M \frac{3 \eta  (\eta +1) }{3 \eta^2+1} \cos \varphi_X
\nonumber\\&&
+\sqrt{M} \left[
-\frac{i \sqrt{3} \sqrt{7\eta^2-4\eta+1}}{2\sqrt{3\eta^2+1}} \sin \varphi_X \left(  b^{\dagger}_X - b^{\phantom{\dagger}}_X \right)
+\frac{\sqrt{3} (\eta-1) (3\eta+1)}{2 \left(3\eta^2+1\right)} \cos \varphi_X \left( c^{\dagger}_X + c^{\phantom{\dagger}}_X \right)
\right],
\\
S_X^y &=& M \frac{3 \eta  (\eta +1)}{3 \eta^2+1} \sin \varphi_X
\nonumber\\&&
 +\sqrt{M}\left[
 \frac{i \sqrt{3} \sqrt{7 \eta^2-4 \eta +1}}{2 \sqrt{3 \eta^2+1}} \cos \varphi_X \left( b^{\dagger}_X- b^{\phantom{\dagger}}_X \right)
+\frac{\sqrt{3} (\eta -1) (3 \eta +1)}{2 \left(3 \eta^2+1\right)} \sin \varphi_X \left( c^{\dagger}_X + c^{\phantom{\dagger}}_X \right)
\right],
\\
S_X^z &=& \sqrt{M} 
\left[
-\frac{\sqrt{3} \eta (\eta +1) }{\sqrt{3 \eta^2+1} \sqrt{7 \eta^2-4 \eta +1}} \left( b^{\dagger}_X + b^{\phantom{\dagger}}_X \right)
+\frac{3  (\eta -1)^2}{2 \sqrt{3 \eta^2+1} \sqrt{7 \eta^2-4 \eta +1}} \left( d^{\dagger}_X + d^{\phantom{\dagger}}_X\right)
\right],
\end{eqnarray}
while the spin quadrupole operators are
\begin{eqnarray}
\left(S_X^z\right)^2 &=& M \frac{3 \left(\eta^2+3\right)}{4 \left(3 \eta^2+1\right)}
+ \frac{2 \sqrt{3}\eta  }{3 \eta^2+1} \sqrt{M} \left( c^{\dagger}_X+c^{\phantom{\dagger}}_X \right) ,
\\
\left(S_X^x\right)^2 -\left(S_X^y\right)^2 &=&  M \frac{6 \eta}{3 \eta^2+1}\cos 2\varphi_X   
\nonumber\\&&
+\left[
-\frac{i \sqrt{3} (\eta +1) (3 \eta -1) }{\sqrt{3 \eta^2+1} \sqrt{7 \eta^2-4 \eta +1}} \sin 2\varphi_X \left(b^{\dagger}_X - b^{\phantom{\dagger}}_X \right)
+\frac{\sqrt{3} \left(3 \eta^2-1\right)}{3 \eta^2+1} \cos 2\varphi_X \left( c^{\dagger}_X + c^{\phantom{\dagger}}_X \right)
  \right. \nonumber\\&& \left.
-\frac{6 i (\eta -1) \eta}{\sqrt{3 \eta^2+1} \sqrt{7 \eta^2-4 \eta+1}} \sin 2\varphi_X \left(d^{\dagger}_X - d^{\phantom{\dagger}}_X\right)
\right] \sqrt{M},
\\
S_X^x S_X^y + S_X^y S_X^x  &=&  M \frac{6 \eta}{3 \eta^2+1}  \sin 2\varphi_X 
   \nonumber\\&&
+\left[
 \frac{i \sqrt{3} (\eta +1) (3 \eta -1)}{\sqrt{3 \eta^2+1} \sqrt{7 \eta^2-4 \eta +1}} \cos 2\varphi_X \left(b^{\dagger}_X - b^{\phantom{\dagger}}_X \right)
+\frac{\sqrt{3} \left(3 \eta^2-1\right)}{3 \eta^2+1} \sin 2\varphi_X \left( c^{\dagger}_X + c^{\phantom{\dagger}}_X \right)
  \right. \nonumber\\&& \left.
+\frac{6 i (\eta -1) \eta}{\sqrt{3 \eta^2+1} \sqrt{7 \eta^2-4 \eta +1}} \cos 2\varphi_X \left(d^{\dagger}_X - d^{\phantom{\dagger}}_X\right)
\right]
   \sqrt{M},
\\
S_X^x S_X^z + S_X^z S_X^x  &=&  \sqrt{M} \left[
-\frac{\sqrt{3} (\eta +1) (3 \eta -1)}{\sqrt{3 \eta^2+1} \sqrt{7 \eta^2-4 \eta +1}} \cos \varphi_X \left(b^{\dagger}_X + b^{\phantom{\dagger}}_X \right)
+i \sqrt{3} \sin \varphi_X \left( c^{\dagger}_X - c^{\phantom{\dagger}}_X \right)
  \right. \nonumber\\&& \left.
-\frac{6 (\eta -1) \eta}{\sqrt{3 \eta^2+1} \sqrt{7 \eta^2-4 \eta +1}} \cos \varphi_X \left(d^{\dagger}_X + d^{\phantom{\dagger}}_X\right)
\right],
\\
S_X^y S_X^z + S_X^z S_X^y  &=&  \sqrt{M} \left[
-\frac{\sqrt{3} (\eta+1) (3 \eta -1)}{\sqrt{3 \eta^2+1} \sqrt{7 \eta^2-4 \eta +1}} \sin \varphi_X \left(b^{\dagger}_X + b^{\phantom{\dagger}}_X \right)
-i \sqrt{3} \cos \varphi_X \left( c^{\dagger}_X - c^{\phantom{\dagger}}_X \right)
  \right. \nonumber\\&& \left.
-\frac{6 (\eta-1) \eta d^{\phantom{\dagger}}_X}{\sqrt{3 \eta^2+1} \sqrt{7 \eta^2-4 \eta +1}}  \sin \varphi_X \left(d^{\dagger}_X + d^{\phantom{\dagger}}_X\right)
\right],
 \end{eqnarray}
\end{subequations}
\end{widetext}
where terms that are proportional to $M$ and $\sqrt{M}$ are shown.

The multiboson spin-wave Hamiltonian up to quadratic order in bosons reads:
\begin{eqnarray}
\mathcal{H}\approx M^2 \mathcal{H}^{(0)}+ M^{3/2} \mathcal{H}^{(1)}+ M \mathcal{H}^{(2)}
\end{eqnarray}
where $\mathcal{H}^{(0)}$ is equal to mean field energy (\ref{eq:SF_en0}), $\mathcal{H}^{(1)}$ is identically zero when (\ref{eq:eta_sol}) is satisfied, and the quadratic term has the following form for the solution given by Eqs.~(\ref{eq:eta_sol}):
\begin{widetext}
\begin{eqnarray}
\mathcal{H}^{(2)} 
& = & 
+\frac{6 (\eta+1)^2 \left(9 \eta ^3-5 \eta^2-\eta +1\right)}{\left(3\eta^2+1\right) \left(7\eta^2-4\eta+1\right)}
 J \left(b_A^{\dagger} b_A + b_B^{\dagger} b_B \right) 
 +\frac{72 \eta ^3 (\eta+1)^2 }{\left(3\eta^2+1\right) \left(7\eta^2-4\eta+1\right)} J 
 \left(d_A^{\dagger} d_A+d_B^{\dagger} d_B\right) \nonumber\\&&
+\frac{9 (\eta-1)^4}{\left(3\eta^2+1\right) \left(7\eta^2-4\eta+1\right)}  J_z
 \left( d_A^{\dagger} d_B^{\dagger} + d_A d_B + d_A^{\dagger} d_B + d_B^{\dagger} d_A \right)
\nonumber\\&&
-\frac{6 \sqrt{3} \eta (\eta+1)  (\eta-1)^2}{\left(3\eta^2+1\right) \left(7\eta^2-4\eta+1\right)} J_z
\left( d_A b_B+ d_B b_A+b_A^{\dagger} d_B + b_A^{\dagger} d_B^{\dagger}+ b_B^{\dagger} d_A+b_B^{\dagger} d_A^{\dagger}+d_A^{\dagger} b_B+d_B^{\dagger} b_A  \right)
\nonumber\\&&
+\frac{36 \sqrt{3} \eta^2 (\eta+1)  (\eta-1)^2}{\left(3\eta^2+1\right) \left(7\eta^2-4\eta+1\right)} J
\left( b_A^{\dagger} d_A +  b_B^{\dagger} d_B + d_A^{\dagger} b_A + d_B^{\dagger} b_B \right)
\nonumber\\&&
+\left[\frac{12  \eta^2 (\eta+1)^2}{\left(3\eta^2+1\right) \left(7\eta^2-4\eta+1\right)} J_z -\frac{3 \left(7\eta^2-4\eta+1\right)}{3 \eta^2+1} J \cos 2\varphi \right] 
\left(b_A^{\dagger} b_B^{\dagger} +  b_A b_B + b_A^{\dagger} b_B + b_B^{\dagger} b_A \right) \nonumber\\&&
+\frac{3 (3\eta+1)(\eta-1) \sqrt{7\eta^2-4\eta+1}}{\left(3\eta^2+1\right)^{3/2}}
i J \sin 2\varphi\left(  
  b_A c_B   
+ b_A c_B^{\dagger}   
+ b_B^{\dagger} c_A  
+ b_B^{\dagger} c_A^{\dagger}
- b_A^{\dagger} c_B  
- b_A^{\dagger} c_B^{\dagger}  
- b_B c_A   
- b_B c_A^{\dagger}   
\right)\nonumber\\&&
+\frac{3 (3\eta+1)^2 (\eta-1)^2 }{\left(3\eta^2+1\right)^2} J \cos 2\varphi
\left( c_A^{\dagger} c_B^{\dagger} +  c_A c_B + c_A^{\dagger} c_B +  c_A c_B^{\dagger} \right)
 +6 J (\eta+1) \left( c_A^{\dagger} c_A + c_B^{\dagger} c_B\right).
\end{eqnarray}
\end{widetext}
We note that in zero field ($\varphi=\pm \pi/2$) and parallel spins the Hamiltonian separates into two parts, one involving $b$ and $d$ bosons, the other only the $c$ bosons. Similarly, for the uniform state in high fields, where the variational parameters are given by Eq.~(\ref{eq:eta_sol_sat}) we get
$\mathcal{H}^{(2)}=\mathcal{H}^{(2)}_{bd} + \mathcal{H}^{(2)}_{c} $ with 
\begin{widetext}
\begin{eqnarray}
\mathcal{H}^{(2)}_{bd} 
& = &
\frac{9 J_z (\eta-1)^4}{\left(3\eta^2+1\right) \left(7\eta^2-4\eta+1\right)}
\left(d_B d_A + d_A^{\dagger} d_B + d_A^{\dagger} d_B^{\dagger} + d_B^{\dagger} d_A \right)
\nonumber\\&&
-\frac{6 \sqrt{3} J_z \eta (\eta+1) d_A b_B (\eta-1)^2}{\left(3\eta^2+1\right) \left(7 \eta^2-4 \eta +1\right)}
\left(
 d_A b_B  
+ d_B b_A 
+ b_A^{\dagger} d_B
+ b_A^{\dagger } d_B^{\dagger} 
+ b_B^{\dagger} d_A  
+ b_B^{\dagger} d_A^{\dagger} 
+ d_A^{\dagger} b_B
+d_B^{\dagger} b_A
\right)
\nonumber\\&&
+\left(
\frac{12 J_z \eta^2 (\eta +1)^2}{\left(3\eta^2+1\right) \left(7\eta^2-4\eta+1\right)}
-\frac{3 J \left(7\eta^2-4\eta+1\right)}{3 \eta^2+1}
\right) 
\left( b_B b_A + b_A^{\dagger} b_B^{\dagger} \right) 
\nonumber\\&&
+\left(
\frac{12 J_z \eta^2 (\eta+1)^2}{\left(3\eta^2+1\right) \left(7\eta^2-4\eta+1\right)}
+\frac{3 J \left(7 \eta^2-4 \eta +1\right)}{3 \eta^2+1}
\right) 
\left( b_A^{\dagger} b_B + b_B^{\dagger } b_A \right) 
\nonumber\\&&
+\left(
\frac{ g_{xx} h_x (\eta+1) \left(9 \eta ^3-5 \eta^2-\eta +1\right)}{2 \eta \left(7\eta^2-4\eta+1\right)}
-\frac{6 J (\eta+1)^2 \left(9 \eta ^3-5 \eta^2-\eta +1\right)}{\left(3 \eta^2+1\right) \left(7\eta^2-4\eta+1\right)}
\right) 
\left( b_A^{\dagger} b_A + b_B^{\dagger} b_B \right) 
\nonumber\\&&
+\left(
\frac{3 \sqrt{3} g_{xx} h_x (\eta-1)^2 \eta }{7\eta^2-4\eta+1}
-\frac{36 \sqrt{3} J (\eta-1)^2 \eta^2 (\eta+1)}{\left(3\eta^2+1\right) \left(7 \eta^2-4 \eta +1\right)}\right)
\left( b_A^{\dagger} d_A+ b_B^{\dagger} d_B + d_A^{\dagger} b_A + d_B^{\dagger} b_B\right)
\nonumber\\&&
+\left(
\frac{6 g_{xx} h_x \eta^2 (\eta+1)}{7\eta^2-4\eta+1}
-\frac{72 J \eta ^3 (\eta+1)^2}{\left(3\eta^2+1\right) \left(7\eta^2-4\eta+1\right)}\right) 
\left(  d_A^{\dagger} d_A + d_B^{\dagger} d_B \right)
\end{eqnarray}
and
\begin{eqnarray}
\mathcal{H}_c^{(2)} 
& = &
+\left(\frac{ g_{xx} h_x \left(3 \eta^2+1\right)}{2 \eta }-6 J (\eta+1)\right) 
\left( c_A^{\dagger} c_A+  c_B^{\dagger} c_B \right)
\nonumber\\&&
+\frac{3 J (3\eta+1)^2 (\eta-1)^2}{\left(3\eta^2+1\right)^2}
\left( c_B c_A + c_A^{\dagger} c_B + c_A^{\dagger} c_B^{\dagger} + c_B^{\dagger} c_A \right).
\end{eqnarray}
\end{widetext}

%-------------------------------
\subsection{The spectrum in zero field}
%-------------------------------

The finite anisotropy reduces the symmetry down to $O(2)$, and the Goldstone mode associated with turning the order parameter in the $xy$ plane is desribed by the 
\begin{eqnarray}
\gamma^\dagger =  \gamma &\propto& 
2 \eta  (\eta +1) \left(
b_A^{\dagger} + b_B^{\dagger} 
+b_A^{\phantom{\dagger}} + b_B^{\phantom{\dagger}} 
\right) \nonumber\\
&&-\sqrt{3} (\eta -1)^2
   \left(
d_A^{\dagger} + d_B^{\dagger} 
+d_A^{\phantom{\dagger}} + d_B^{\phantom{\dagger}} 
   \right)
\end{eqnarray}
so that $[\mathcal{H}^{(2)},\gamma^\dagger]=[\mathcal{H}^{(2)},\gamma]=0$.

The energy of the remaining 5 modes:
In the $d_-$ mode the spin moves in the $xy$ plane and is of the same symmetry character as the Goldstone mode, with energy
\begin{equation}
 \omega_{d_-} = \frac{18 (\eta+1) \sqrt{
\eta \left(\eta^3-\eta^2+3 \eta +1\right)}}{ 3 \eta^2+1 } J.
\end{equation}
The energy of the $b_+$ and $d_+$ modes that depend on the $j_z=J_z/J$ is given by the
\begin{widetext}
\begin{eqnarray}
\frac{\omega^4}{J^4} =
  \left(\frac{6\eta +6}{3 \eta^2+1}\right)^2 \left(9 \eta^4 +9 \eta^3 +17 \eta^2 + 7 \eta +2 -8 \eta^2 j_z\right) \frac{\omega^2}{J^2}
- 72 \eta^3 \left(\frac{6\eta +6}{3 \eta^2+1}\right)^4 \left(4 \eta^3 -\eta^3 j_z+\eta^2 j_z-3 \eta j_z-j_z\right).
\end{eqnarray}
\end{widetext}
equation. The four aforementioned modes consist of $b$ and $d$ bosons.

 Finally, the two ``stretching'' modes that involve the $c$ bosons only:
\begin{eqnarray}
\omega_{c_-} &=& 
\frac{6 \sqrt{(\eta+1) \left(9 \eta^5+18 \eta^4-6
 \eta^3+4 \eta^2+5 \eta +2\right)}}{3 \eta^2+1} J
\nonumber\\&&
\\
 \omega_{c_+} &=&
\frac{6 \sqrt{\eta (\eta+1) \left(9 \eta^4+18 \eta^2+8 \eta -3\right)}}{3 \eta^2+1} J
\end{eqnarray} 

In the limit of small $\Lambda$ the energies are
\begin{subequations}
\begin{eqnarray}
 \omega_{b_-} &=& 0,\\
 \omega_{b_+}^2 &=& 24 J \left[ 3 (J - J_z) + \Lambda\right] + O(\Lambda^2), \\
 \omega_{c_-} &=& 12 J + \Lambda + \frac{1}{8}\frac{\Lambda^2}{J} + O(\Lambda^3/J^2), \\
 \omega_{c_+} &=& 12 J + \Lambda - \frac{1}{24}\frac{\Lambda^2}{J} + O(\Lambda^3/J^2), \\
 \omega_{d_-} &=& 18 J + O(\Lambda^3/J^2), \\
 \omega_{d_+} &=& 18 J + O(\Lambda^3/J^2). 
\end{eqnarray}
\end{subequations}
The $\omega_{b_+}$ shows the typical square-root behaviour of the anisotropy gap on the exchange anysotropy $J-J_z$ and single-ion anisotropy $\Lambda$ for small gap.

In the limit of large single-ion anisotropy ($\Lambda\gg J,J_z$):
\begin{subequations}
\begin{eqnarray}
 \omega_{b_-} &=& 0,\\
 \omega_{b_+}^2 &=&  32 J (4 J -J_z + 12 J^2/\Lambda) +O(1/\Lambda^2), \\
 \omega_{c_-} &=& 2 \Lambda + 7 J + O(J^2/\Lambda), \\
 \omega_{c_+} &=& 2 \Lambda + J + O(J^2/\Lambda), \\
 \omega_{d_-} &=& 2 \Lambda + J + O(J^2/\Lambda), \\
 \omega_{d_+} &=& 2 \Lambda + 7 J+ O(J^2/\Lambda). 
\end{eqnarray}
\end{subequations}

\end{document}